\documentclass[a4paper]{article}

\usepackage{amsmath,amsfonts,epsfig,graphics,amssymb}
\usepackage{hyperref}
\usepackage{authblk}

\title{\bf Calibration of “Troitsk nu-mass” detector readout electronics by signal digital filters}
\author[1,2]{S.\,B.~Abdiganieva}
\author[1]{A.\,I.~Berlev}
\author[1,2]{M.\,A.~Bochkov}
\author[1]{N.\,A.~Likhovid}
\author[1,2]{V.\,S.~Pantuev}
\author[1]{S.\,V.~Zadorozhny}

\affil[1]{\it Institute for Nuclear Research RAS, 117312 Moscow, Russia}
\affil[2]{\it Moscow Institute of Physics and Technology, Dolgoprudny, Moscow region 141700, Russia}

\date{}

\begin{document}
%\maketitle

%\twocolumn[
%\begin{@twocolumnfalse}
\maketitle
\begin{abstract}
{We present the results of tuning and calibration of the detector electronics in the signal digitization mode. The goal of the experiment is to search for a possible sterile neutrino signature in tritium beta-decay. The read-out electronics work in direct oscilloscope mode, which requires to optimize time frame the with the goal to minimize noise and energy resolution. We use a 7-pixel silicon drift detector (SDD) and a CMOS charge sensitive preamplifier with very low integration capacitor. Amplifier forms a slowly rising output shape and operates in pulse-reset mode. The 125 MHz ADC digitizes the signals. Using calibration data from $Fe^{55}$ and $Am^{241}$ gamma sources we check triangular and trapezoid digital filters to obtain the best noise and energy resolution performance. We are also examining the option to differentiate the output signal.}
\vspace{\baselineskip}
\end{abstract}
%\end{@twocolumnfalse}]

%%% PACS numbers
%\PACS{74.50.+r, 74.80.Fp}

%\vspace{\baselineskip}

\paragraph{1. Introduction.}
Nowadays many physics experiments rely on high tech achievements and solutions in detector and electronics performance. In the “Troitsk nu-mass” setup we use a 7-pixel silicon drift detector (SDD)~\cite{Lechner} with direct digitization electronics. The goal of “Troitsk nu-mass” is to search for a new hypothetical particle – sterile neutrino~\cite{Abdurashitov:2015jha}. If this particle exists, it could close a long-standing problem of the particle Standard Model: where are the right-handed neutrinos, could sterile neutrinos be such particles? In our experiment, we make precise measurements of the electron energy spectrum from the tritium beta-decay in attempt to find a change in the spectrum shape. The “Troitsk nu-mass” setup consists of a gaseous tritium source and an electrostatic spectrometer~\cite{Belesev:2013cba} with a silicon drift detector for electron registration. In our setup the signals from the detector are continuosly digitized, and the task is to perform triggering, fast signal sampling anf filtering. The details of the detector readout are the subject of this work. We describe several  digital filter applications that allow us to decrease noise, trigger threshold and to improve off-line amplitude resolution.

\paragraph{2. Detector and electronics.}

\begin{figure}[htb]
	\begin{center}
		\includegraphics[width=.8\linewidth]{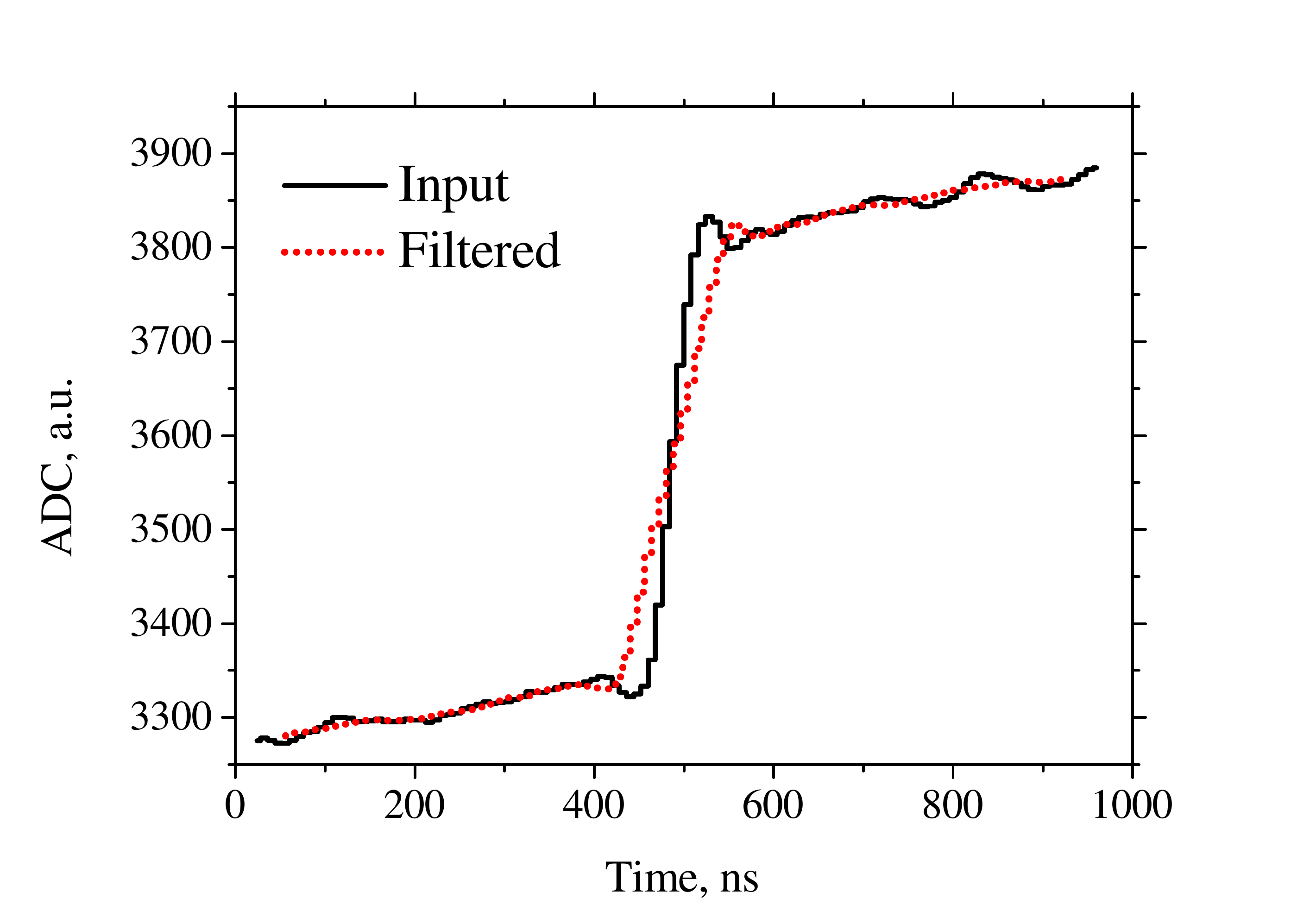}
	\end{center}
	\caption{Linear rising output signal with a step-like signature from the registered particle.  Dotted line is the result of moving average oven 16 samples. }
	\label{fig:Kard_Sergey}
\end{figure}

As mentioned, we use a silicon drift detector (SDD) with 7 pixels 2 mm each for detection of electrons with energies up to 20 keV. The major advantage of such a detector is very a low anode capacitance, less than 100 fF. The detector was manufactured at the Semiconductor Laboratory of the Max Planck Society (HLL)~\cite{Lechner}. Each anode is bonded to a charge-sensitive amplifier, CUBE~\cite{link}, which operates in pulse-reset mode, with a capacitor value of 20 fF in the signal feedback loop.  CUBE is a CMOS preamplifier for radiation detectors. The read-out system was manufactured by XGLab – Bruker Nano Analytics~\cite{link}. It is an 8-channel CUBE Bias Board, XGL-CBB-8CH. 
Similar detectors were already tested at the “Troitsk nu-mass” spectrometer in 2017-2019~\cite{tristan} and demonstrated excellent  energy resolution, less than 200 eV at full width at half maximum (FWHM).
Continues signal digitization is provided by a 16-channel TQDC module in the VME case designed in the Joint Institute for Nuclear Research, JINR, Dubna, for the NICA project~\cite{NICA}. The TQDC module allows one to operate both in continuos digitization mode and with FPGA sub-processing. Each TQDC channel has 12 bit ADC at a sampling rate of 125 MHz. Here, as mentioned, we discuss the applicability in continuous mode and the usage of an external reset mode for front-end electronics. 
	The amplifier output signal is continuously linearly rising from -0.5~V to +0.5~V because of the unavoidable detector leak current and  
 is reset by external signal with time period of 100-200 microseconds. 
In the case of a particle signal, its charge integrates and produces a step-like change in the linear rising output signal, 
Fig.~\ref{fig:Kard_Sergey}.  

\paragraph{3. Signal sampling and filtering.}

\paragraph{ 3.1	Trigger threshold.}
During actual physics measurements of the tritium energy beta-spectrum, one of the critical parameters is the electronics threshold applied. The lower this value, the more precise the data. To optimize the performance of the read-out system, we should use a digital moving average threshold. For this purpose, we can apply, for example,  the simplest linear transformation or moving averaging like:
\begin{equation}
	a_i = \frac{\sum_{i-\frac{m}{2}}^{i+\frac{m}{2}-1}x_i}{m},
\label{eq:1}	
\end{equation}
where $a_i$ is moving average,  $x_i$ is ADC value in $i-$th bin, $m$ is interval of averaging. 

We collect a set of calibration data with $Fe^{55}$ and $Am^{241}$ isotopes in direct digitization mode with a ADC time step of 8 nsec. Dotted line in Fig.~\ref{fig:Kard_Sergey} illustrates how the moving average works. For each trigger time window is set so that
 the real signal step is in the middle of the frame. Then, we apply the ADC averaging over some number of time bins, $m$. One can see that the noise fluctuations are significantly suppressed; some signal oscillations around the rising edge have almost completely diappeared. The signal amplitude restoration is significantly improved. 

The above example could be used to set the threshold if it were a  “normal” signal over a {\it constant} bias value. To perform a fine-tuning of our digital threshold over a continuously rising signal we have to do a slightly different transformation, namely, to apply the well-known triangular digital filter, $a_i={\hat{H}\cdot \hat{x}}$, where $\hat{H}=(-m, m)$ denotes some filter coefficients. For the triangular filter it sets  “-1” and “+1” repeated $m$-times each. This filter sums over the bin interval size $m$, then subtracts the sum over the {\it previous} $m$ bins, and normalizes  the difference by $m$. As soon as the filter slides over a continuously rising signal with a constant slope, the output of the filter would be a constant. Approaching the step region, it finds the step and gives a triangular shaped digital signal with a rising and falling side equal to $m$. Automatically, the noise will be suppressed at some level. In the case of a trapezoidal filter, $\hat{H}=(-m,l, m)$ , where $l$ denotes some number or $zeros$ between subtracting intervals and defines the width of the trapezoid top.

\begin{figure}[htb]
	\begin{center}
		\includegraphics[width=.8\linewidth]{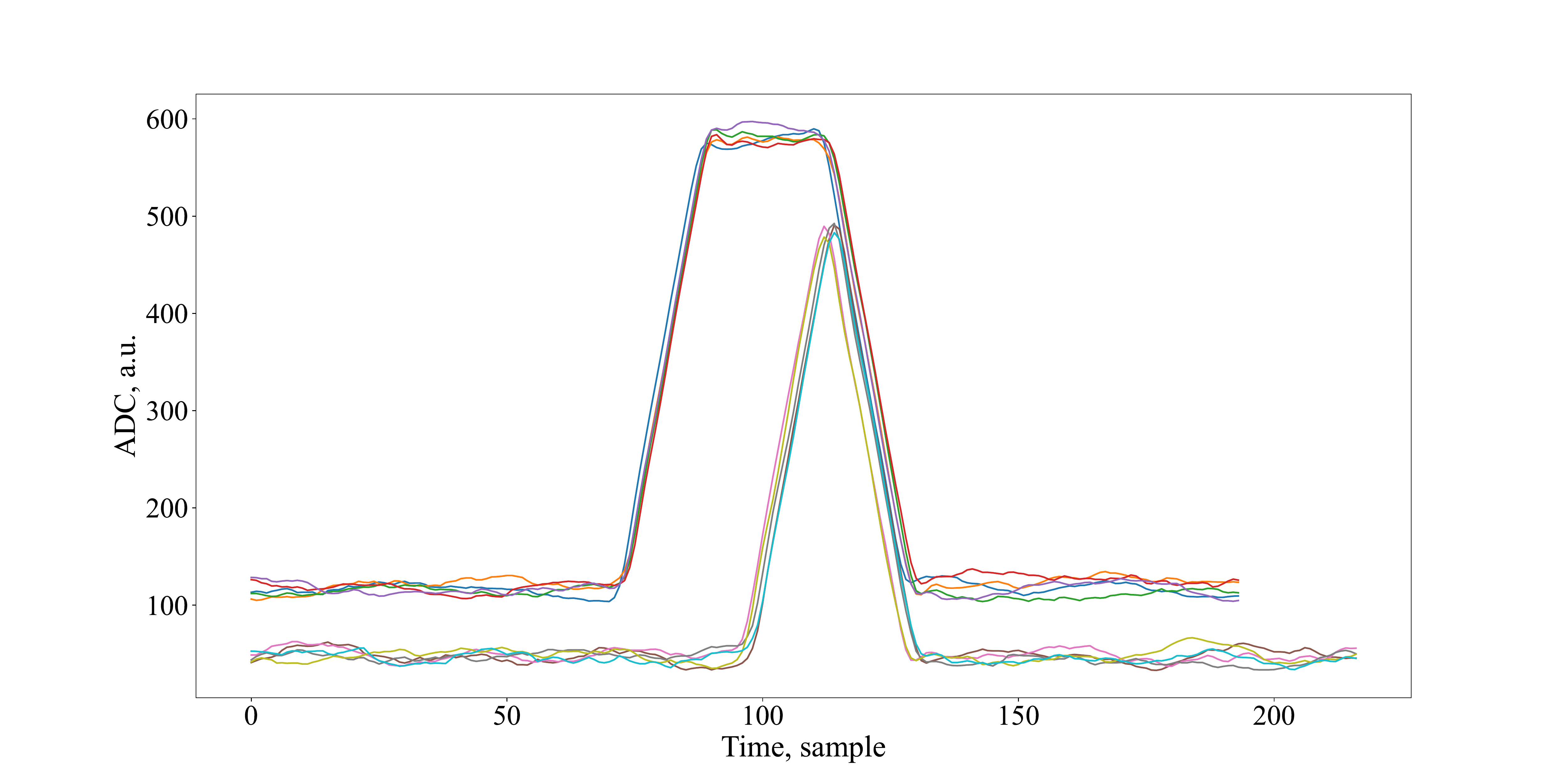}
	\end{center}
	\caption{Results of application of trapezoidal and triangular filters for a several events with $m$=16~bins and $l$=16~bins for the $Fe^{55}$ source.  }
	\label{fig:Kard}
\end{figure}

Fig.~\ref{fig:Kard} demonstrates the result of application of the triangular and trapezoid filters with $m$=16 bins and $l$=16 bins to the real data. Pay attention, that for a continuously rising signal the digital “zero energy” or noise line changes, depending on the parameters $m$ and $l$.

\begin{figure}[htb]
	\begin{center}
		\includegraphics[width=.8\linewidth]{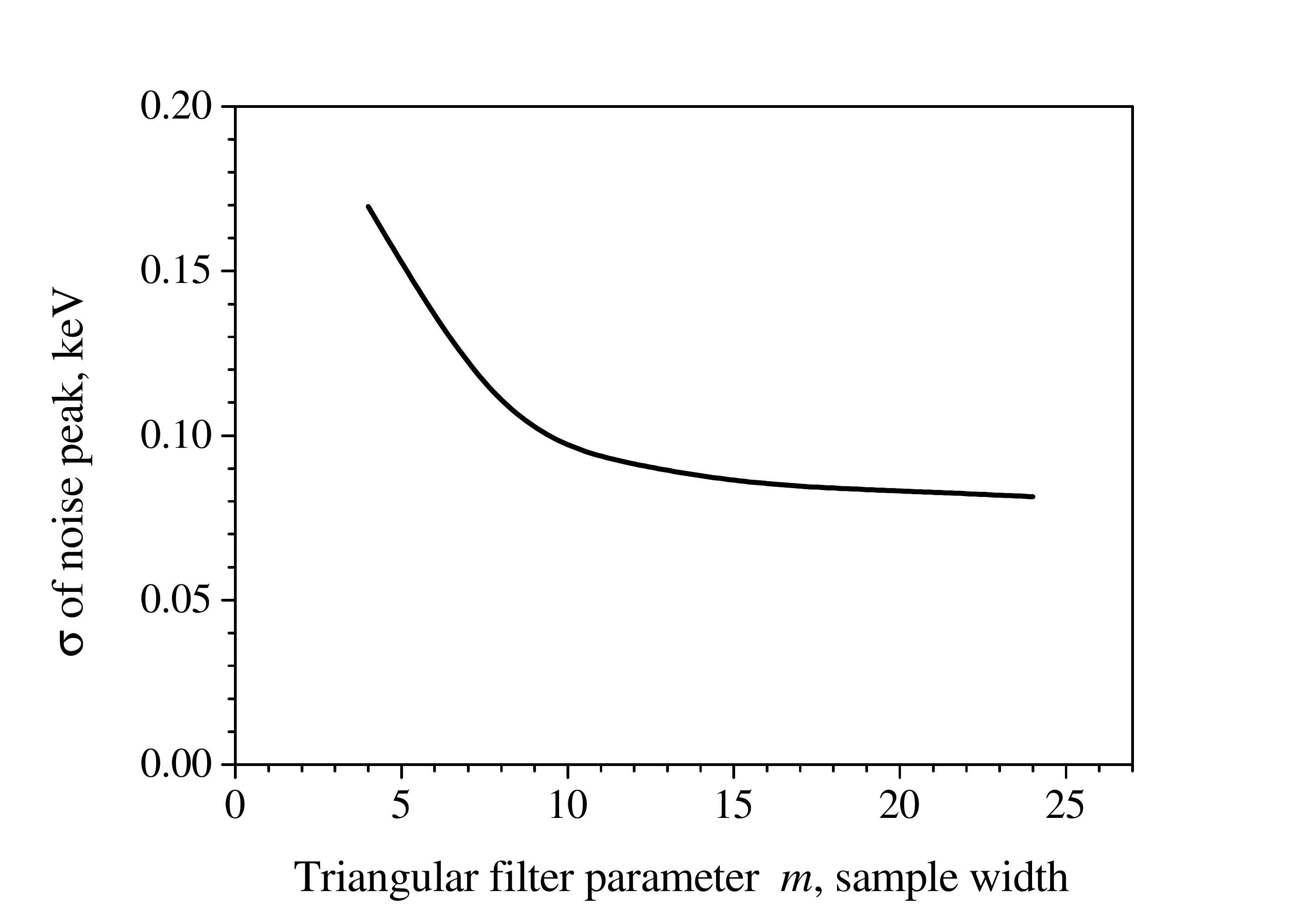}
	\end{center}
	\caption{The noise width ($\sigma$) versus $m$ for triangular filtering.}
	\label{fig:pila_noise}
\end{figure}

To tune the digital threshold to its optimal value, we did a scan of the noise amplitude with a triangular filter for different $m$. In each case, we fit the noise width by the Gaussian function. Fig.~\ref{fig:pila_noise} show dependence of the noise level in keV for different triangular filter parameter $m$. The noise sharply drops at $m>$ 8 bins and then almost saturates at $m > $16. For $m$=16 we get noise level at 3$\sigma$ as 0.27 keV,  which estimates a possible digital threshold . 
We checked the noise dependence for the trapezoidal filter for different $l$ at $m$= 16. We don't get any improvement in the noise level by increasing the parameter $l$ from 0 to 24 bins. Thus, we conclude that triangular filter works well for noise suppression. 

\begin{figure}[htb]
	\begin{center}
		\includegraphics[width=.8\linewidth]{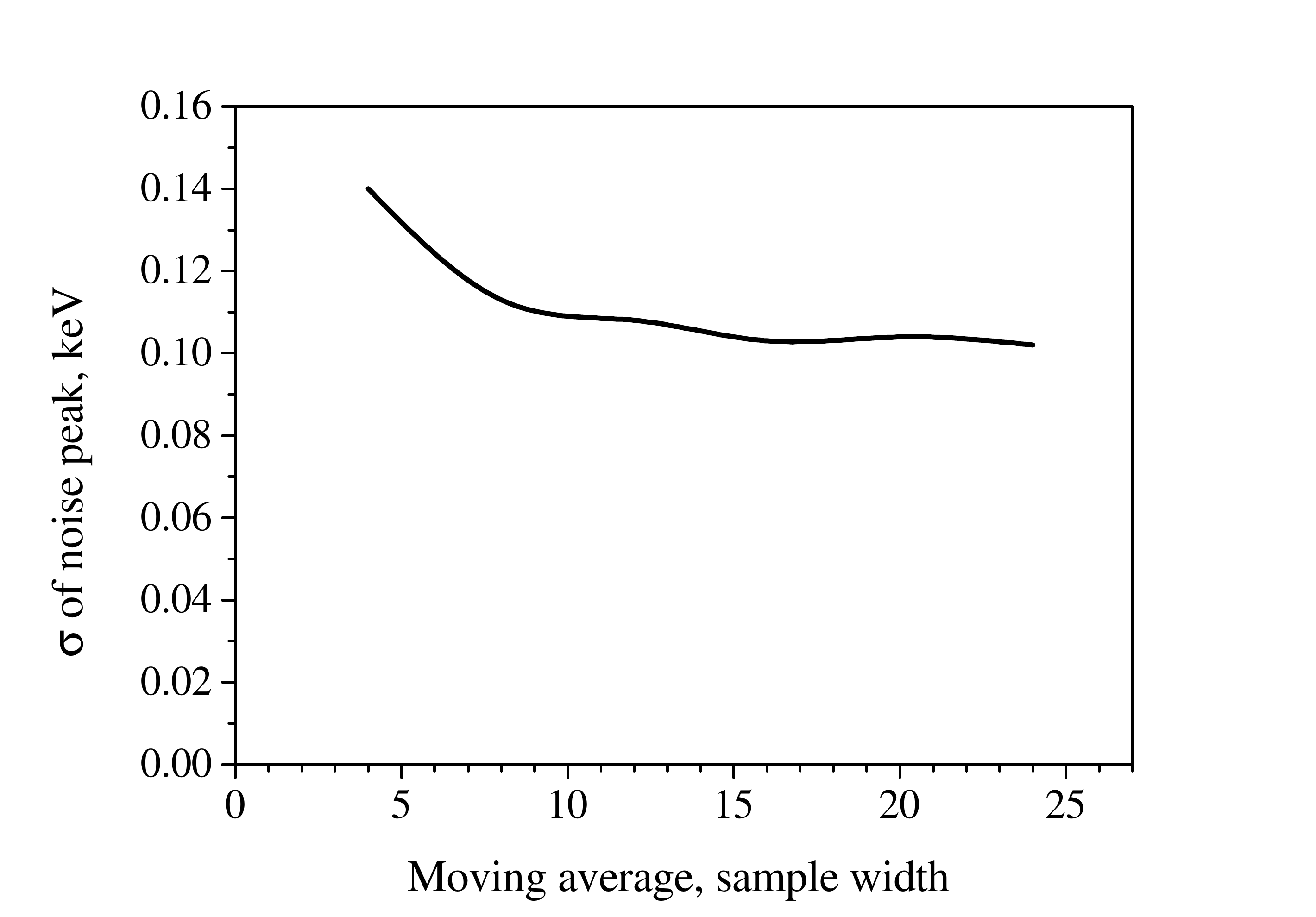}
	\end{center}
	\caption{The noise level for the differentiated signal versus the moving average filter sample width, $m$.   }
	\label{fig:diff_noise}
\end{figure}

We also use a configuration when a continuously rising signal is differentiated by a decoupled 1~nF capacitor.  In this case, for trigger selection we can use a standard moving average filter with a normalized sum over some number of bins. Fig.~\ref{fig:diff_noise} demonstrates the dependence of the noise level on the sample width. The best value is worse by about 20\% relative to the triangular filter in Fig.~\ref{fig:pila_noise}.

\begin{figure}[htb]
	\begin{center}
		\includegraphics[width=.8\linewidth]{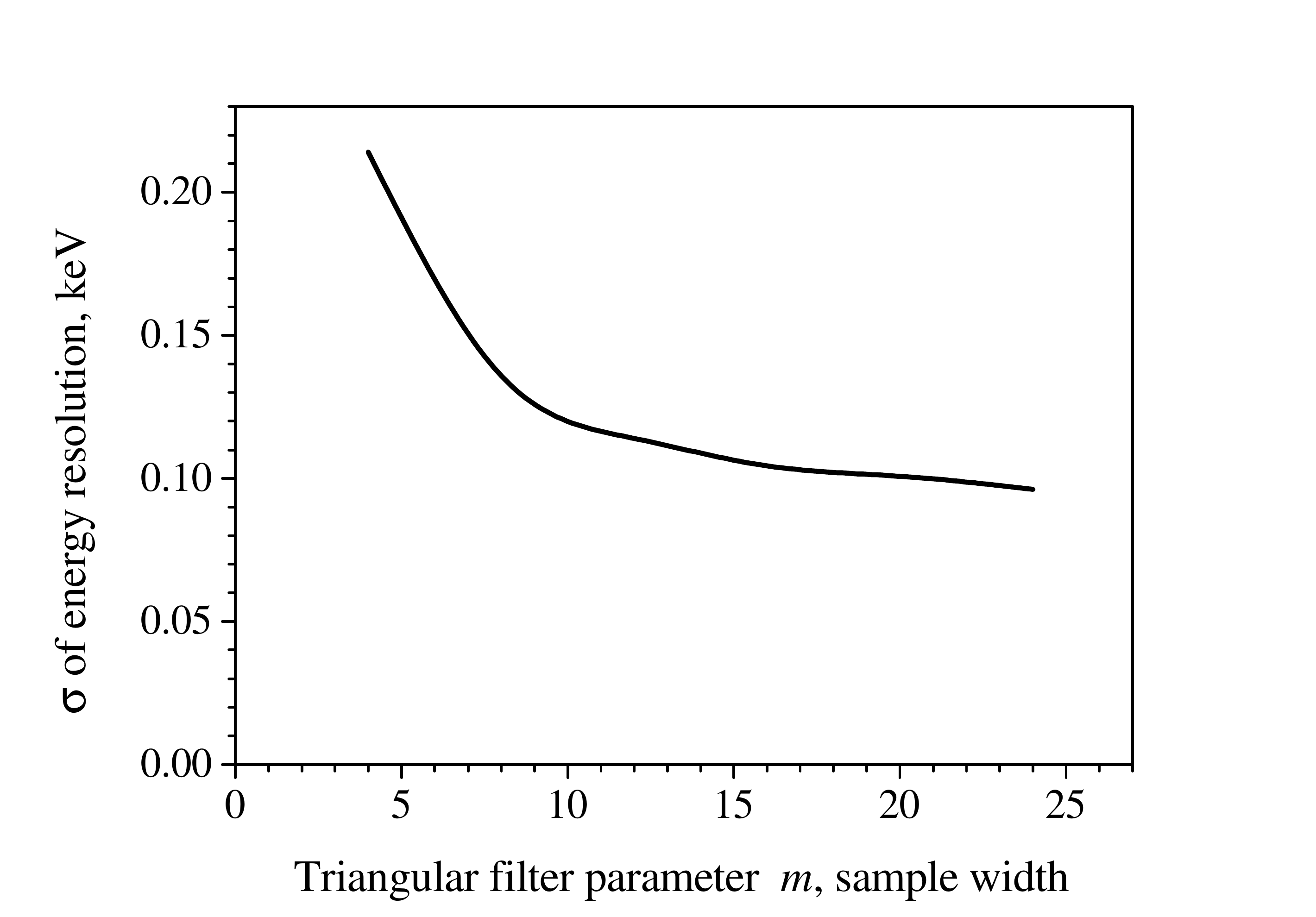}
	\end{center}
	\caption{Energy resolution ($\sigma$) of the 5.9 keV $Fe^{55}$ peak versus the triangular filter parameter $m$.  }
	\label{fig:pila_sigma}
\end{figure}

\begin{figure}[htb]
	\begin{center}
		\includegraphics[width=.8\linewidth]{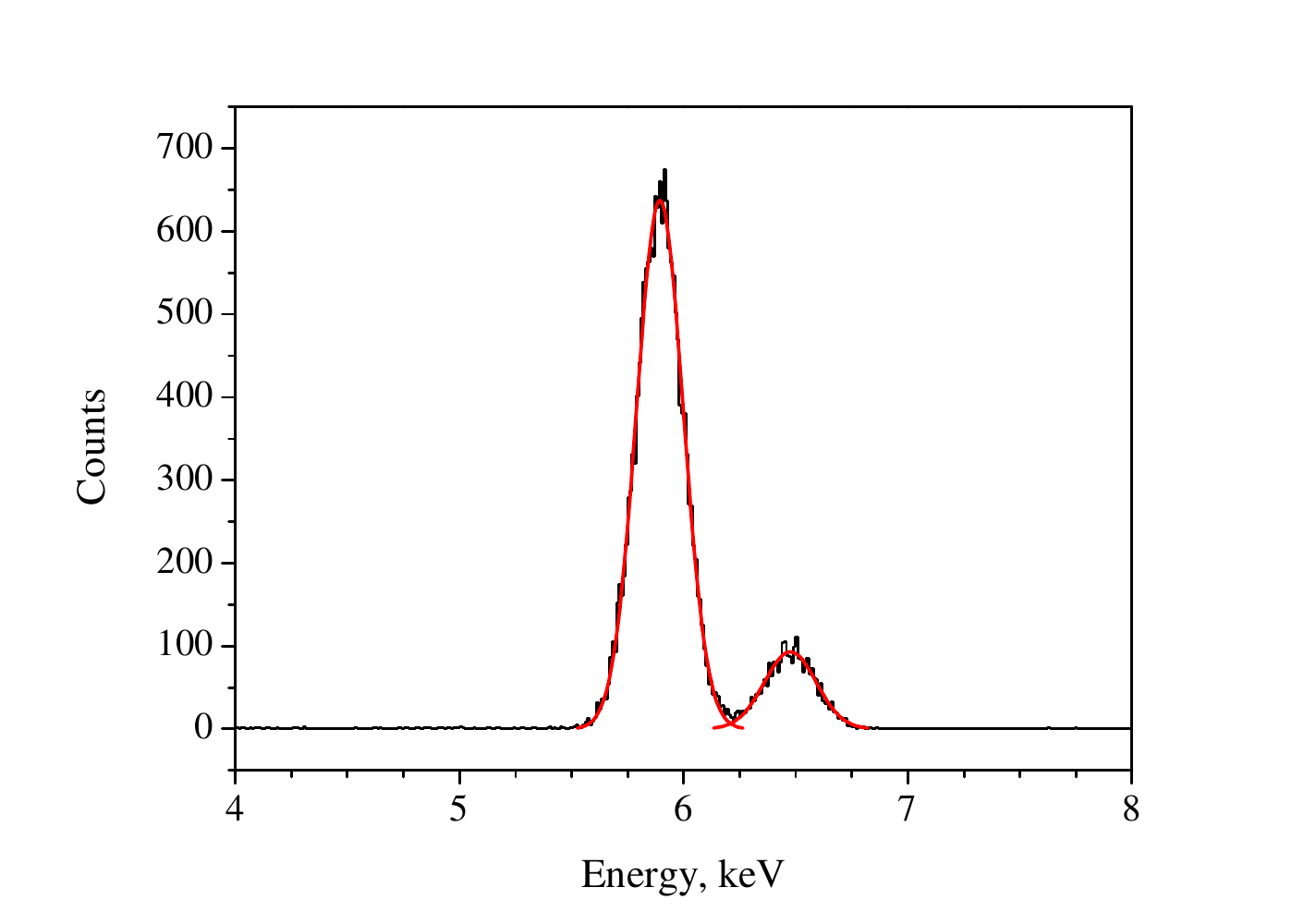}
	\end{center}
	\caption{$Fe^{55}$ gamma spectrum by triangular filter at $m$=16 and fit with $\sigma$=104~eV. One can clearly see the main  peak at 5.89~keV and the smaller one at 6.49~keV.    }
	\label{fig:Fe55_spectrum}
\end{figure}

\paragraph{ 3.2 Optimization of energy resolution}
For each trigger, we need to optimize the length of the time window (sample width) with a goal to keep it short at a reasonable energy resolution. We collect calibration data with a $Fe^{55}$ gamma  source. The source has an intense gamma line at 5.89~keV and a smaller one at 6.49~keV. The time frame was set to be wide, about two microseconds with a step-like signal in the middle. We apply triangular and trapezoidal filters to restore the signal amplitude. 
Fig.~\ref{fig:pila_sigma} shows the energy resolution after applying the triangular filter. Again, one can see that the energy resolution saturates at $m$ above 16. The energy spectrum of $Fe^{55}$ source is shown in Fig.~\ref{fig:Fe55_spectrum} at $m$=16. It is appropriate to mention that the detector energy resolution $\sigma$=0.11 keV is worse the best value for SDD detector of FWHM = 150~eV (or $\sigma$ about 0.062 keV) as published in~\cite{resol} . However, that value was obtained at SDD – 20C temperature , and we also know from our experience~\cite{tristan} that this detector has a larger leakage current compared to a similar detectors we used before. 

\begin{figure}[htb]
	\begin{center}
		\includegraphics[width=.8\linewidth]{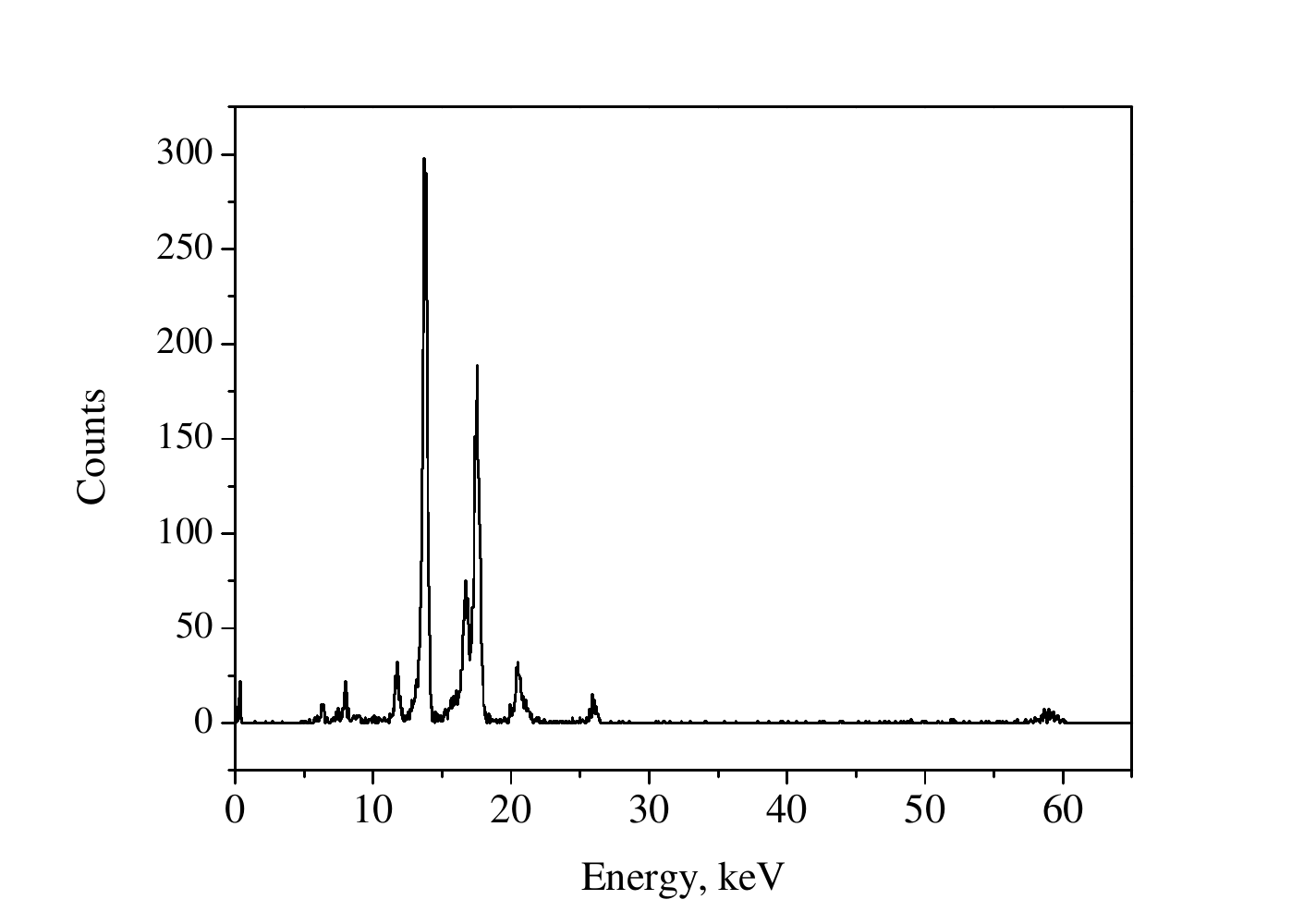}
	\end{center}
	\caption{$Am^{241}$ gamma spectrum, triangular filter at $m$=16.}
	\label{fig:Am241_spectrum}
\end{figure}

 We also take data with the $Am^{241}$ gamma source. Fig.~\ref{fig:Am241_spectrum}  shows the spectrum reconstructed with a triangular filter at $m$=16. One can see prompt 13-18~keV lines and a low intensity line from the metastable decay at 59.54 keV.
Applying trapezoidal filter at $m$=16 and at different $l$,  within the error the energy resolution, has no improvement compared to the triangular (or $l$=0) filter result.  Thus, we conclude that the usage of a triangular filter at $m$=16 bins will satisfy our needs for threshold selection and energy resolution. The minimal time window for the ADC sampling should be $=16 +16\cdot 2=48$ bins.

\begin{figure}[htb]
	\begin{center}
		\includegraphics[width=.8\linewidth]{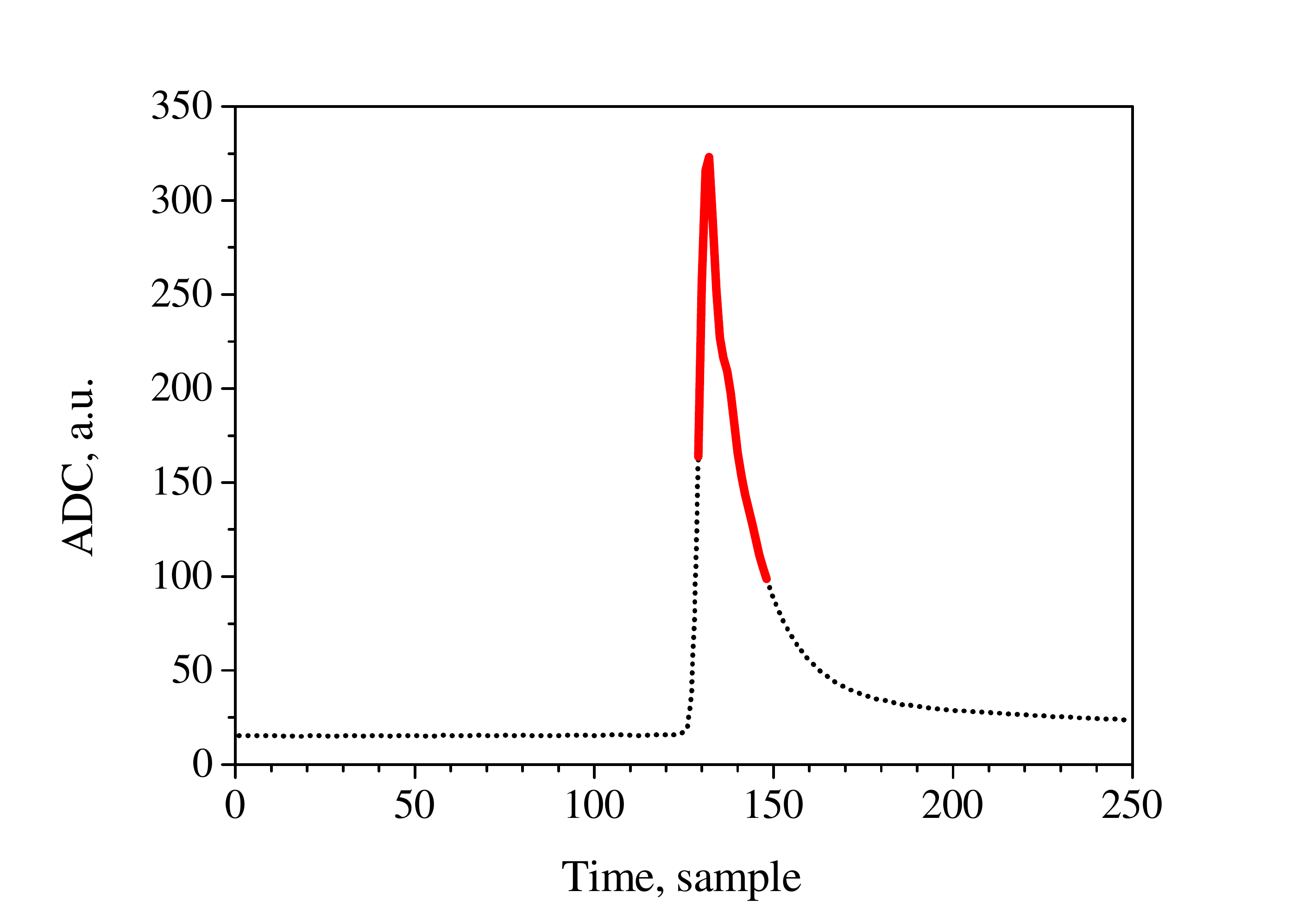}
	\end{center}
	\caption{The shape of the differentiated signal (dotted line), and the optimum range to get the best energy resolution (solid line).    }
	\label{fig:kadr_Lih}
\end{figure}

\paragraph{3.3	Energy resolution for a differentiated signal}
The elementary particle releases in the silicon detector some portion of charge, which forms the output signal. In the current detector, charge integration requires about 50~nsec or a maximum of~80 nsec. This time defines the signal rise time from the bias board. In some applications, it is more convenient to differentiate a linear rising output from the bias board to get the “normal”, not liear rising,  signal. The charge collection time fluctuates between events, any passive filter, like a differentiation in our case, will distort the amplitude, and a “ballistic deficit” occurs~\cite{ballistic}, which is the loss of output signal amplitude due to the interplay between the finite charge-collection times in a detector and the characteristic time constants of the electronics.   In  Fig.~\ref{fig:kadr_Lih} we show the digitized signal from $Fe^{55}$ with a dotted line. The shape was averaged over many events. To get the best energy resolution and to minimize the ballistic deficit we try a few methods. By taking just maximum in the frame, we get the energy resolution of $\sigma$=187.1~eV. Another approach is to integrate in the certain neighbourhood of the waveform maximum. The ADC sum from the maximum by -8 bins to the left and 30 bins to the right gives $\sigma$ = 132~eV.  Varying the ranges to the left and to the right from the maximum, we find the best resolution of 116.6~eV for -3 samples on the left and 19 samples on the right (combination of -3,19). Another method is to introduce weights for the ADC values. We check a simple one, where the weight is proportional to the value itself, that is, we calculate $\sum x_i\cdot*x_i / \sum x_i$. By taking the whole rise time range, -8 bins, and 30 bins to the right from the maximum, we get a resolution of 126~eV. Varying the range, the best resolution is again for combination (-3, 19) and $\sigma$ = 116.4~eV, which is the same as for a simple sum in this range, and about 10\% worse than for a trianglular filter on liner rising signal, Fig.~\ref{fig:pila_sigma}. In case of the differentiated signal, the minimal time window for the ADC sampling should also be around 50 bins.
\paragraph{4. Conclusions.}
We investigate the application of digital filters for ``Troitsk NuMass'' signal processing. A silicon drift detector with 7 pixels 2~mm each is used for electron and gamma detection in the energy range under~20 keV. The goal was to optimize the trigger parameters and the time window for digitization at a clock frequebcy of 125~MHz  in direct ADC oscilloscope mode. Calibration data with $Fe^{55}$ and $Am^{241}$ gamma isotopes were analyzed by triangular and trapezoid digital filters. The triangular filter with integration over 16 ADC samples gives the optimal noise and energy resolution. The best energy resolution for a gamma line of 5.9~keV is 110~eV (sigma). Application of a trapezoidal filter with a flat top up to 24~bins (or 200~nsec) does not improve this result. It was concluded that the triangular filter with an integration range of 16~bins will satisfy our needs for trigger selection and energy resolution. We also checked the detector performance with a differentiated output signal.  The simple moving average filter minimizes the noise level at integration over 12-16~samples. A few methods were tested to get the best energy resolution. We found that a simple integration over a rather narrow interval around the ADC maximum gives an energy resolution of about 116 eV for differentiated signal. The minimal time window for the ADC sampling should be around 50 bins.

\paragraph{Acknowledgements.}
We would like to thank our colleges from Max Planck Institute of Physics in Munchen and, particularly, Susanne Mertens, for their help at the beginning of this work.  The work was supported by the Ministry of Science and Higher Education of the Russian Federation under Contract 075-15-2020-778.

\end{document}